\newcommand{\be}{\begin{equation}}
\newcommand{\ee}{\end{equation}}
\newcommand{\bea}{\begin{eqnarray}}
\newcommand{\eea}{\end{eqnarray}}
\newcommand{\ba}[1]{\begin{array}{#1}}
\newcommand{\ea}{\end{array}}

\newcommand{\diracslash}[1]{#1\llap{/\kern2pt}}

\documentclass[aps,pra,twocolumn,showpacs,amsmath,amssymb]{revtex4}
\usepackage{epsfig}
\usepackage{amssymb}
\usepackage{times}
\usepackage{amssymb}
\usepackage{amsmath}
\usepackage{mathrsfs}
\usepackage{graphicx}
\usepackage{epsfig}
\usepackage{dcolumn}
\usepackage{color}
\usepackage{bm}
\begin{document}
\title{Resonant Manipulation of  $d$-wave Interaction of Cold Atoms with Two Lasers and a Magnetic Field }
\author{Bimalendu Deb}
\affiliation{Department of Materials Science,
Raman Center for Atomic, Molecular and Optical Sciences,
Indian Association
for the Cultivation of Science,
Jadavpur, Kolkata 700032, India.}

\begin{abstract}
We present a theory for manipulation of $d$-wave interaction of cold atoms with  two lasers strongly driving  two  photoassociative
transitions.  The theory predicts the occurrence  of a coherence between two excited
ro-vibrational bound states due to the  photoassociative dipole-couplings of ground-state $d$-wave scattering state
to the bound states. We show that this excited-state coherence  significantly influences atom-atom  interaction.  In particular, 
this leads to the enhancement of $d$-wave elastic scattering  and to  the suppression of  inelastic scattering. 
In the presence of an $s$-wave magnetic Feshbach resonance, the two lasers can couple
the $s$-wave resonance  with the  $d$-wave scattering state leading to
the further  enhancement in $d$-wave scattering at relatively low energy.
 Our numerical calculations based on realistic parameters
show that $d$-wave manipulation would be most effective in case of atoms having excited diatomic states with  narrow natural
linewidth. We estimate that at 100 $\mu$K  the inelastic scattering rate in Yb can be reduced to 20 s$^{-1}$ while the elastic scattering rate
can be two orders of magnitude larger.
\end{abstract}


\maketitle

\section{introduction}

Cold atoms have now become a testing ground for interacting many-particle physics, 
in particular for models of condensed matter systems.
They  offer a unique opportunity to study many-body physics \cite{bloch}  with a controllable two-body interaction.  
By making use of the tunability of the  $s$-wave scattering length with an external magnetic field, $s$-wave physics 
of interacting  ultracold atoms has been extensively studied in recent times. To go beyond the $s$-wave physics, 
it is essential to devise methods for manipulation of higher partial-wave interactions of cold atoms.  
Higher partial-wave interactions are important for some exotic phases of matter such as
the $p$-wave anisotropic superfluidity of liquid He-3 and the  $d$-wave  superconductivity of  cuprate materials
\cite{htsup}. 
The ability to study phenomena related to  $d$-wave Cooper pairing  with fermionic atoms will help us to gain new insight 
into the open problem of high temperature superconductivity. 
To explore $d$-wave pairing  of fermions with two spin components (like electrons)  at various interaction regimes,  
it is of primary  interest to devise an effective method of manipulating the  $d$-wave interaction
in two-component fermionic atoms.

There are mainly two methods of
controlling atom-atom  interaction at ultracold temperatures. The most
widely used method involves the   magnetic Feshbach resonance (MFR)
\cite{feshbach,revfesh}. The
other method using the  optical Feshbach resonance (OFR) \cite{fedichev,
ofrexpt} is a relatively recent one.  Over more than a decade, the  MFR  has been
 used to study the   $s$-wave physics of interacting atomic gases.
In fact, the MFR has become   a standard tool for studying Fermi superfluidity \cite{ketterle} and  
strongly interacting Fermi gases \cite{fermi}
with tunable  $s$-wave scattering length.
The OFR has been originally  developed as a method for altering the $s$-wave scattering length as in the MFR.
This method relies on off-resonant free-bound photoassociative  transitions.
The use of strong  photoassociative coupling  has been proposed as an optical method of
manipulating  higher partial-wave interactions \cite{debprl}. Of late,  $p$-wave OFR has been experimentally 
reported in fermionic $^{171}$Yb atoms by Yamazaki {\it et al.} \cite{yamazaki} based on  a  
suggestion by Goyal {\it et al.} \cite{pra:2010:goel} to use 
purely long-range states \cite{prl:2008:takahashi,pra:2009:deutsch} of photoassociated excited molecules. 
Earlier, magnetic field induced $p$-wave Feshbach resonances have been observed in spin-polarized 
$^{40}$K \cite{pwave1} and $^6$Li \cite{pwave2} Fermi gases.  
Not only $p$-wave OFR, but alo generating other higher partial-wave interactions in ultracold atoms under various physical conditions  
has attracted  a great deal of research  interest \cite{spielman,prl2009levinsen,nishida,pra2011gianna,laburthe} in recent times. 
The generation of $d$-wave interaction in ultracold atoms by strong confinement has been discussed in  Refs \cite{nishida,pra2011gianna}.     
However, an efficient manipulation of the  $d$-wave  atom-atom interaction for the purpose of studying 
$d$-wave many-body physics with cold atoms is yet to be demonstrated.  Both $p$- and  $d$-wave 
scattering amplitudes of cold atoms are in general extremely small according to Wigner threshold laws. 
 This 
poses a major challenge  in manipulating $p$- and  $d$-wave interactions with an optical or any other field. The method 
we propose  here is applicable for both $p$- and $d$-wave manipulation, but in this work we specifically 
concentrate on $d$-wave only.

\begin{figure}
\includegraphics[width=3.5in]{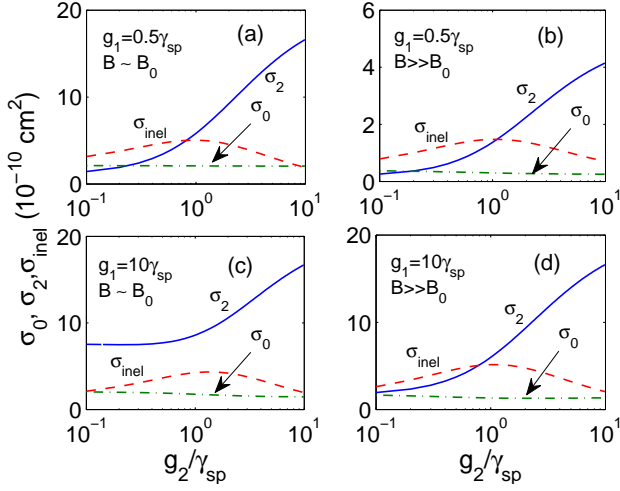}
\caption{Plots of  $s$- and $d$-wave elastic scattering cross sections $\sigma_0$ (dashed-dotted curves) 
and $\sigma_2$ (solid curves), respectively,
 and total  inelastic scattering cross section $\sigma_{inel}$ (dashed curve ) for $^6$Li atoms 
 as a function of
dimensionless scaled intensity ($g_2/\gamma_{sp}$) of laser 2 for different values of the intensity
of laser 1 and magnetic field: (a) $g_1 = 0.5 \gamma_{sp}$ and $(B-B_0)/|\Delta| = -0.24$, 
( where $\Delta$ is the width of the MFR)
(b) $g_1 = 0.5 \gamma_{sp}$ and $B\gg B_0$, (c)  $g_1 = 10 \gamma_{sp}$ and $(B-B_0)/\Delta = -0.24$, and 
(d) $g_1 = 10  \gamma_{sp}$ and $B \gg B_0$. $\gamma_{sp}$ is assumed to be 2 MHz and  energy $E = 20$  $\mu$K. }
 \label{fig2}
  \end{figure}

Here we present a theory for an all optical or magneto-optical manipulation of $d$-wave interaction in three
dimensions.
Consider two colliding ground-state cold atoms irradiated by
two  photoassociation (PA) lasers tuned near resonance with two excited ro-vibrational bound (molecular) 
states 1 and 2
characterized by two different rotational quantum numbers $J_1$ and $J_2$, respectively. However,
these states may have the same or different vibrational quantum numbers.
These ro-vibrational states must be accessible by PA transitions from the  $d$-wave scattering state of the two ground-state atoms.
 The method we propose here relies on the quantum interference between the two PA transition pathways leading to the occurrence of
a coherence  between the two excited states.
 We show that this coherence causes  suppression of inelastic scattering and enhancement of $d$-wave elastic scattering. 
Note that
the inelastic scattering considered here is due to absorption from the PA lasers and loss due to spontaneous emission.
Because of spontaneous emission from excited states, one-color OFR
is bound to suffer from inelastic loss  as discussed by Bohn and Julienne \cite{bj97}. However, here we show 
that, by establishing a coherence between two excited ro-vibrational levels by two lasers, it is possible to suppress
this inelastic loss significantly.  Previous studies  \cite{jpb:2009:gsa,jpb:2010:deb,jpb:2009:arpita} on Fano-Feshbach  resonances  in  PA    \cite{jpb:2009:gsa}
in the presence of a MFR \cite{prl1998heinzen,prl1999vladan,prl:2008:junker}  primarily focused on controlling $s$-wave scattering states,
 suppression of inelastic scattering \cite{jpb:2010:deb} and control over power broadening \cite{jpb:2009:arpita}.
The present work is basically concerned with optical manipulation of the  anisotropic atom-atom interaction. 
In case of magnetic atoms,  a magnetic field may  be applied along the $z$ axis
to induce an $s$-wave MFR. For two-component fermionic atoms or spin-polarized bosonic atoms, it is possible 
to optically couple the  $s$-wave MFR to  the $d$-wave scattering state by exploiting appropriate rotational selection rules 
for the two photoassociative transitions as analyzed below. 
The results of our numerical calculations based on realistic parameters  reveal that the optical manipulation
of the  $d$-wave interaction is most effective  in atoms having
narrow natural linewidths such as Yb.

\section{Anisotropic dressed continuum}
Here we present a dressed state description of the ground-state  continuum interacting with a magnetic field 
and 
a pair of laser fields.

\subsection{The model and its analytical solution}

For simplicity,  let us consider only two ground-state scattering channels $\mid g_1\rangle $ and   $\mid g_2\rangle $ 
with  $\mid g_1\rangle $ being open (continuum) and
$\mid g_2\rangle $ closed. To keep our discussions most general at the outset, we here do not 
 specify the compositions of the channels, but we will give details  of the channels in a specific  numerical illustration 
 later. The closed channel is assumed to support an $s$-wave bound state $\mid \chi \rangle $ whose binding energy lies close 
to the threshold of the  $\mid g_1\rangle $ channel.   These two ground-state channel thresholds and the 
closed-channel bound state
may be tunable with an external magnetic field leading to an $s$-wave Feshbach resonance. The ground-state channels 
are asymptotically 
diagonal. However, at intermediate separations there 
is an interchannel coupling $V(r)$ which depends on the interatomic separation $r$. 

The energy spacings between the excited bound states $\mid b_1 \rangle$ and $\mid b_2 \rangle$ 
characterized essentially by the two different rotational quantum 
 numbers $J_1$ and $J_2$  are assumed to be large enough 
 so that the two lasers $L_1$ and $L_2$  can drive only these two  bound levels, respectively. 
 If these two bound states belong to the same vibrational level $v$, then the rotational energy spacing between them 
 must be large enough. Furthermore, these bound levels are assumed to be far below the respective dissociation threshold 
 so that the transition
 probability at the single-atom level is negligible. The excited potentials asymptotically correlate to two separated atoms with 
 one being in the electronic ground $S$ state and another in the excited $P$ state. If the two atoms are homonuclear, then the 
 excited state potentials asymptotically go as $C_3/r^3$ where $C_3$ represents the  resonant dipole-dipole interaction coefficient. 
 In the case of heteronuclear atoms, the excited-state potential goes as $\sim \, - 1/r^6$ at large separations. 
 
In the rotating wave approximation, the Hamiltonian of these three
bound states interacting with the ground continuum can be expressed as $\hat{H} = \hat{H}_0 + \hat{H}_I$ where 
\bea
 \hat{H}_0 &=&  \sum_{n} (E_{n} - \hbar \omega_{L_n}) \mid b_n \rangle \langle b_n \mid \nonumber \\
&+& E_{\chi} \mid \chi \rangle \langle \chi \mid \otimes \mid g_2 \rangle \langle g_2 
+ \sum_{\ell m_{\ell}} \int E' dE' \nonumber \\
&\times& \mid E' \ell m_{\ell} \rangle_{\rm{bare}} \, _{\rm{bare}} \langle E' \ell m_{\ell} \mid \otimes \mid g_1 \rangle \langle g_1 \mid 
\eea
is the free part of the Hamiltonian with $E_n$ representing  
the binding energy of $n$-th excited molecular state $\mid b_n \rangle$, 
$E_{\chi}$ being the energy of the cosed-channel bound state $\mid \chi \rangle$ and 
$\mid E',\ell m_{\ell} \rangle_{\rm{bare}}$ being the  bare (unperturbed) partial-wave
($\ell m_{\ell}$)  scattering state with energy  $E'$. Note that all the energies are measured from the 
open-channel threshold unless stated otherwise. The interaction part of the Hamiltonian is 
\bea 
\hat{H}_I &=& \sum_{n,M_n} \sum_{\ell m_{\ell}} \int d E' \Lambda_{J_n M_n}^{\ell m_{\ell}}(E')  
\mid b_n \rangle \langle E' \ell m_{\ell} \mid \langle g_1 \mid  \nonumber \\ &+& 
\int d E' V_{\chi 0}(E') \mid \chi \rangle \, _{\rm{bare}}\langle E' 0 0 \mid \otimes \mid g_2 \rangle \langle g_1 \mid 
\nonumber \\
&+& \sum_{n,M_n} \Omega_{n\chi} \mid b_n \rangle \langle \chi \mid \langle g_2 \mid  + \rm{c.c.}  
\eea 
Here $ \Lambda_{\ell m_{\ell}}^{J_n M_n}(E)$ is the dipole
matrix element of  transition $  \mid b_n \rangle  \rightarrow  \mid E, \ell m_{\ell} \rangle_{\rm{bare}}\mid g_1 \rangle$, 
$V_{\chi 0}(E)$ is the coupling between the quasibound ($\chi$) state and the $s$-wave ($\ell = 0$) scattering 
state of the  bare continuum
and $\Omega_{n\chi}$ is the Rabi frequency between the $n$th excited bound state and the $\chi$ state.

Let  $\mid E,{\hat{k}} \rangle$ represent an eigen ket that satisfies the time-independent Schr\"{o}dinger 
equation
$\hat{H} \mid E,{\hat{k}} \rangle = E \mid E,{\hat{k}} \rangle$, where 
${\hat{k}}$ denotes a unit vector along the incident relative momentum of the two atoms. These eigen kets are 
energy normalized in a manner such that $\langle E' \hat{k}' \mid E, \hat{k} \rangle = \delta(E - E')
\delta(\hat{k} - \hat{k}')$.  To solve 
the Schr\"{o}dinger equation, we expand  $\mid E,{\hat{k}} \rangle$ in the following form 
\bea
&&\mid E,{\hat{k}} \rangle = \sum_{n=1,2}\sum_{M_n}  A_{n E}
 \mid b_n(J_n,M_n) \rangle +  B_{E} \mid \chi \rangle \mid g_2 \rangle \nonumber \\
&+& \sum_{\ell,m_{\ell}} \sum_{\ell',m_{\ell'}}  \int d E' C_{E',\ell m_{\ell}}^{\ell' m_{\ell'}} Y_{\ell' m_{\ell'}}^{*}(\hat{k})\mid E',\ell m_{\ell} \rangle_{\rm{bare}}
\mid g_1 \rangle \nonumber \\
\label{eq1} \eea
where $B_E$,  $A_{nE}$ and  $C_{E',\ell m_{\ell}}^{\ell' m_{\ell'}}$ 
are the expansion coefficients to be derived.   The coefficients $B_E$ and  $A_{nE}$ 
can be expanded as  $B_E = \sum_{\ell' m_{\ell'}} B_{E}^{\ell' m_{\ell'}} Y_{\ell' m_{\ell'}}^{*}(\hat{k})$
and  $A_{n E} =
\sum_{\ell' m_{\ell'}} A_{n E}^{\ell' m_{\ell'}} Y_{\ell' m_{\ell'}}^{*}(\hat{k}) $. Physically,  
$A_{n E}^{\ell' m_{\ell'}}$  
implies the probability amplitude for the excitation of the bound state $\mid b_n\rangle$ when  
the incident partial-wave of relative motion of the two atoms is $\ell'$ and its projection
along the space-fixed $z$ axis is $m_{\ell'}$. Similarly, $B_{E}^{\ell' m_{\ell'}}$ 
denotes  the probability amplitude for the occupation of the quasibound state $\mid \chi \rangle$ 
for the $(\ell' m_{\ell'})$ incident partial wave.  The analytical expressions of
the coefficients   $ A_{n E}^{\ell' m_{\ell'}}$, $B_{E}^{\ell' m_{\ell'}}$ and $C_{E',\ell m_{\ell}}^{\ell' m_{\ell'}}$ 
are derived  in Appendix A. They are expressed in terms of the basic  coupling parameters of the model, 
namely the laser couplings $\Lambda_{J_n M_n}^{\ell m_{\ell}}$ and magnetic coupling $V_{\chi 0}$. Explicitly,  
\bea
 A_{nE}^{\ell'm_{\ell'}}(M_n) =
 \frac{\mathscr{A}_{n0}\delta_{\ell',0} 
 + \mathscr{F}_{n\ell'} (1 - \delta_{\ell' 0}) }{ \mathscr{E}_n + i \hbar (\mathscr{G}_n + \gamma_n)/2},
 \label{ae} \eea 
The first term  $\mathscr{A}_{n0}$ in the numerator of Eq. (\ref{ae}) is  defined by Eq. (\ref{a0})
in Appendix A. 
This term arises from incident partial wave $\ell'=0$ ($s$ wave) only. It consists of two terms. The first term
$\beta_{n \epsilon} \Lambda^{0 0}_{J_n M_n}$  is the amplitude for transition form the $s$-wave  
scattering state to the  $n$th bound state in the presence of  a MFR and 
 laser $L_n$. The effect of the 
MFR is given by the dimensionless parameter 
$\beta_{n\epsilon} =  \frac{ q_{nf} + \epsilon}{\epsilon + i}  $.  The parameter $q_{nf}$ is the  Fano-Feshbach asymmetry 
parameter \cite{jpb:2009:arpita}. $\epsilon$ 
is a dimensionless energy parameter defined by Eq. (\ref{epsn}) and related to the MFR phase shift 
$\eta_{\rm{res}}$ by the relation $\cot \eta_{res} = - \epsilon$. In the limit 
$\epsilon \rightarrow \pm \infty$ or equivalently $\eta_{res} \rightarrow 0$ or the coupling 
$V_{\chi 0} \rightarrow 0$ we have $\beta_n \rightarrow 1$.  The second term  
$\frac{1} {\xi}_{n'}  {\cal K}_{nn'} \beta_{n'\epsilon} \Lambda^{0 0}_{J_{n'} M_{n'}}$ 
on the right-hand side (RHS) of Eq. (\ref{a0})
describes the combined effects  of the other ($n'$th) bound state, 
two lasers and the  MFR on the transition amplitude from the  $s$-wave scattering state to the $n$-th ($\ne n'$)  
bound state. The quantity  ${\cal K}_{nn'} = {\cal K}_{nn'}^{\rm{mf}} + {\cal K}_{nn'}^{\rm{LL}}$, where 
${\cal K}_{nn'}^{\rm{mf}}$ defined by  Eq. (\ref{knnpmf}) in Appendix A is an effective coupling 
between the two excited bound states induced by the two lasers in the presence of the MFR provided 
both the bound 
states are accessible by PA transitions from the $s$-wave scattering state. ${\cal K}_{nn'}^{\rm{mf}}$ depends on 
the two Fano-Feshbach asymmetry parameters $q_{1f}$ and $q_{2f}$ and therefore on the
optical dipole transitions from the closed channel $s$-wave quasibound state to the two excited states. 
${\cal K}_{nn'}^{\rm{LL}}$ 
given by Eq. (\ref{knnpll}) in  Appendix A represents an effective coupling between the two excited states 
 induced by the two lasers  through continuum-bound PA transitions. Note that 
${\cal K}_{nn'}^{\rm{LL}}$ depends on the properties of the open-channel bare continuum and  
does not depend on closed-channel quasibound state or the  bound-bound transition.    
All possible incident partial waves  which are allowed by selection rules for 
both the PA transitions contribute to  ${\cal K}_{nn'}^{\rm{LL}}$.  The second term  $\mathscr{F}_{n\ell'}$  in the  numerator
of Eq. (\ref{ae}) is  given by Eq. (\ref{f1}) in Appendix-A. This term describes transition amplitude 
from a nonzero incident partial wave ($\ell' >0$) to the $n$th bound state in the presence of both the lasers
and the MFR. In the denominator of Eq. (\ref{ae}),
$\mathscr{E}_n = E - \hbar \delta_{n} - E_{n}^{\rm{shift}} - \mathscr{E}_{n}^{\rm{mf}} - \mathscr{E}_{n n'}$, where 
$E_{n}^{\rm{shift}}$ is the light shift of $n$-th bound state due to $L_n$ laser only. 
$\mathscr{E}_{n}^{\rm{mf}}$  given by Eq. (\ref{shiftmf}) in  Appendix A is a shift that 
strongly depends on the MFR, in the limit 
$\epsilon \rightarrow \pm \infty$ this shift vanishes.  $\mathscr{E}_{n n'}$ defined 
by Eq. (\ref{shiftLL}) in the Appendix-A  depends  on both  lasers through the coupling ${\cal K}_{n n'}$. 
The total linewidth for the transition to the $n$th bound state is $\mathscr{G}_n + \gamma_n$ where 
$\gamma_n$ is the spontaneous linewidth and 
$\mathscr{G}_n = \Gamma_{n00}^{\rm{mf}} + \sum_{\ell >0, m_{\ell}} \Gamma_{n\ell m_{\ell}}
 +\Gamma_{n n'}$ depends on the magnetic and laser fields. Here 
 $\Gamma_{n00}^{\rm{mf}} = |\beta_{n\epsilon}|^2 \Gamma_{n00}$ which, in the limit
 $\epsilon \rightarrow \pm \infty$ reduces to the $s$-wave partial stimulated linewidth 
$\Gamma_{n00}$. Thus in the absence of the  MFR, the first two terms in $\mathscr{G}_n$ can be 
added to yield   $\Gamma_n = \sum_{\ell, m_{\ell}} \Gamma_{n\ell m_{\ell}}$ which 
is the stimulated linewidth of the $n$th bound state due to  laser $L_n$. 
 $ \Gamma_{n n'}$ is defined by Eq. (\ref{gnn'}) in  Appendix A and explicitly depends on 
 both  lasers through  ${\cal K}_{n n'}$.
 
 Finally, 
the remaining  expansion coefficients of Eq. (\ref{eq1}), the detailed derivation of 
which is given in the Appendix-A, can be expressed as  
\bea 
B_E^{\ell'm_{\ell'}} &=&  \frac{2}{(\epsilon + i) \Gamma_f } \left 
[   V_{\chi 0}(E) \delta_{\ell' 0} + \sum_{n,M_n}  (q_{nf} - i) \right. \nonumber \\
&\times& \left.   \pi \Lambda^{J_n M_n}_{00} V_{\chi 0}(E) 
A_{nE}^{\ell' m_{\ell'}} \right ], \label{bme}
\eea
\bea 
C_{E' \ell m_{\ell}}^{\ell' m_{\ell'}}(E) &=& \delta (E-E') \delta_{\ell \ell'} \delta_{m_{\ell} m_{\ell'}} +
\frac{F^{\ell' m_{\ell'}}_{\ell m_{\ell}}(E,E')}{E - E'} \nonumber \\ \label{ce}
\eea
where 
\bea
F^{\ell' m_{\ell'}}_{\ell m_{\ell}}(E,E')&=& B_E^{\ell' m_{\ell'}} V_{\chi 0}(E')\delta_{\ell ,0} \nonumber \\
&+&
\sum_{n,M_n}   A_{nE}^{\ell' m_{\ell'}} \Lambda_{\ell m_{\ell}}^{J_n M_n}(E'). \label{fll} \eea
Equations (\ref{ae}), (\ref{bme}), (\ref{ce}) and (\ref{fll}) clearly show that 
all the expansion coefficients of Eq. (\ref{eq1})  can be evaluated in terms of the known input  
parameters of our model. 
The effects of the two lasers and the magnetic field on atom-atom scattering are given by  Eq. (\ref{fll})
as discussed in the next section.

\subsection{Elastic and inelastic scattering}

Making use of the analytical solution as outlined above, we now derive scattering $T$ and $S$ matrices. 
The asymptotic form of the dressed continuum wave function
$\langle \mathbf{r} \mid  E, \hat{k}\rangle = \Psi_E(\hat{k}, \mathbf{r}) $, 
where $\mathbf{r}$ denotes relative coordinate,
is given by $\Psi_{E}( \hat{k}, 
\mathbf{r} \rightarrow \infty) \sim  r^{-1} \sum_{\ell m_{\ell}} \psi_{\ell m_{\ell}}(\hat{k}, r) Y_{\ell m_{\ell}}(\hat{r}) $ 
where
\bea & &\psi_{\ell m_{\ell}}( \hat{k}, r) \sim   \sum_{\ell' m_{\ell'}} \left [ \sin\left (k r - \frac{\ell' \pi }{2} + \eta_{\ell'}^{bg} \right ) \delta_{\ell
\ell'} \delta_{m_{\ell} m_{\ell'}} \right.  \nonumber \\ &-&
\left. \pi  F^{\ell' m_{\ell'}}_{\ell m_{\ell}}(E,E') Y_{\ell'm_{\ell'}}^*(\hat{k}) \exp \left [ i (k r - \frac{\ell \pi }{2} + \eta_{\ell}^{bg}) \right ]
\right ], \label{asymp} \eea
with $\eta_{\ell'}^{bg}$ being the background phase shift in the absence of interchannel and laser couplings.
Now, equating the asymptotic form of $\Psi_E(\hat{k}, \mathbf{r})$  with the boundary condition  $ r^{-1} \sum_{\ell m_{\ell}} [  \sin(k r - \ell \pi /2 + \eta_{\ell} )
- T_{\ell m_{\ell}}(\hat{k}) e^{i (k r - \ell \pi /2 + \eta_{\ell}) }] Y_{\ell m_{\ell}}(\hat{r})   $
we  deduce the anisotropic scattering $T$-matrix element $T_{\ell m_{\ell}}(\hat{k}) = \sum_{\ell' m_{\ell'}} T_{\ell m_{\ell},\ell' m_{\ell'}}(E) Y_{\ell' m_{\ell'}}^*(\hat{k})$
where
\bea T_{\ell m_{\ell},\ell' m_{\ell'}}(E) = T_{\ell'}^{bg} \delta_{\ell
\ell'} \delta_{m_{\ell} m_{\ell'}} + \pi F^{\ell' m_{\ell'}}_{\ell m_{\ell}}(E,E) e^{2 i \eta_{\ell}^{bg}} \nonumber \\ 
\label{tmat} \eea
with $T_{\ell}^{bg} = - e^{i \eta_{\ell}^{bg} } \sin(\eta_{\ell}^{bg})$ being the $T$-matrix element in the absence of all  laser and magnetic couplings.
The $S$-matrix element is obtained from $S_{\ell m_{\ell}}(\hat{k}) =
Y_{\ell m_{\ell}}(\hat{k}) - 2 i T_{ \ell m_{\ell}}(\hat{k})$.
The total  scattering cross section can be obtained by means
of the optical theorem as $\sigma_{tot} =  - \frac{4\pi g_s }{k^2}  \sum_{\ell m_{\ell}} {\rm Im} [T_{\ell m_{\ell}, \ell m_{\ell}}]$,
 where the symbol ${\rm Im}$ stands for the imaginary part and $g_s=1(2)$ for two distinguishable (indistinguishable) atoms. The  total elastic scattering
cross section  $\sigma_{el} =  \sum_{\ell} \sigma_{\ell}$ where
$ \sigma_{\ell} =
\sum_{ m_{\ell}} \sigma_{\ell m_{\ell}}$ and
$ \sigma_{\ell m_{\ell}} =
\frac{4\pi g_s}{k^2} \sum_{\ell' m_{\ell'}} |T_{\ell m_{\ell}, \ell' m_{\ell'}}|^2 $.  The total inelastic scattering
cross section  $\sigma_{inel} = \sigma_{tot} - \sigma_{el} $. Making use of Eq. (\ref{tmat}), we have
\bea
\sigma_{inel} = - \frac{4 \pi g_s }{k^2} \sum_{\ell m_{\ell}} \left [  \Im \left  \{  \pi  F_{\ell m_{\ell}}^{\ell m_{\ell}} \right  \}
+   \sum_{\ell', m_{\ell'}}
|\pi F_{\ell m_{\ell}}^{\ell' m_{\ell'}}|^2 \right ]
\label{inel}
\eea
For both the  spontaneous linewidths $\gamma_1 \rightarrow 0$ and $\gamma_2 \rightarrow 0$, we have $\sigma_{inel} \rightarrow 0$ and  the $S$ matrix is unitary,
that is $\sum_{\ell' m_{\ell'}} S_{\ell m_{\ell},\ell' m_{\ell'}}  S_{\ell' m_{\ell'},\ell'' m_{\ell''}}^{*} = \sum_{\ell' m_{\ell'}} S_{\ell m_{\ell},\ell' m_{\ell'}}^* S_{\ell' m_{\ell'},\ell'' m_{\ell''}} = \delta_{\ell \ell''} \delta_{m_{\ell} m_{\ell''}}$.

The first term on the RHS of Eq.(\ref{tmat}) is the isotropic part that  does not depend on 
laser and magnetic fields. The second term proportional to $ F^{\ell' m_{\ell'}}_{\ell m_{\ell}}(E,E)$ 
contains all the effects of laser and magnetic fields on scattering. The term 
 $F^{\ell' m_{\ell'}}_{\ell m_{\ell}}(E,E)$  given by Eq. (\ref{fll}) is a sum of three terms corresponding  
 to the three dressed bound-state amplitudes. 
For both  lasers to influence $p$- or $d$-wave scattering, it is essential that the  
rotational quantum numbers of the two excited bound states should be such that 
both of them are allowed by selection rules for PA transitions from the  $p$- or $d$-wave ground scattering state.
Once bound state 1 or 2 is populated 
or their coherent superposition is formed, a nonzero partial-wave interaction in the ground continuum can be generated 
by two possible stimulated bound-free transition pathways. 
These pathways can interfere constructively under appropriate conditions
resulting in enhanced higher partial-wave scattering amplitude. The sum of the two terms in 
$F^{\ell' m_{\ell'}}_{\ell m_{\ell}}(E,E)$ for $\ell \ne 0$ amounts to a coherent superposition 
of the amplitudes of those two transitions pathways. Furthermore, when  a nonzero partial-wave ground 
scattering state is strongly coupled simultaneously to a pair of excited rovibrational states, a coherence
between the two excited states due to the coupling $\mathscr{K}_{12}$ is generated. In the next section, we discuss
 how this  coherence  influences ground-state collisional properties.

\section{Results and discussions}

\begin{figure}
\includegraphics[width=3.5in]{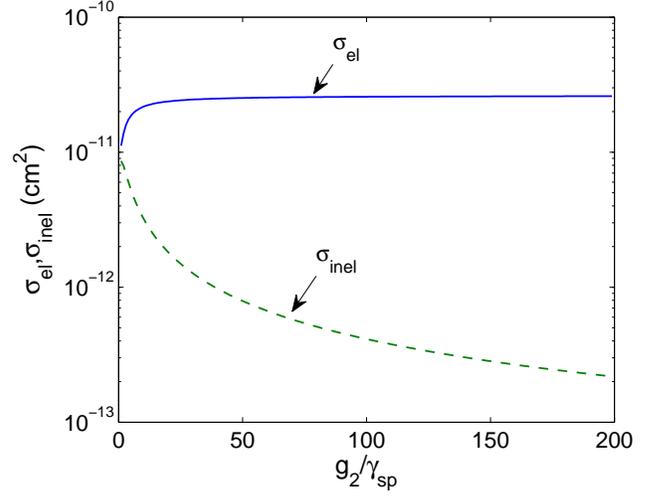}
\caption{This figure is for $^{174}$Yb atoms. Plotted are  the total elastic scattering cross section $\sigma_{el}$ (solid) and total
inelastic scattering cross section $\sigma_{inel}$ (dashed) as a function of $g_2/\gamma_{sp}$ for $g_1 = 200 \gamma_{sp}$ at $E = 100 \mu$K. The value
of $\gamma_{sp}$ is 200 kHz.
}
 \label{fig4}
  \end{figure}

The analysis presented 
in the preceding section is quite general since it is applicable to  any ultracold atomic gas. 
Equation (\ref{eq1}) 
represents a nonperturbative eigenvalue solution of the model and is expressed as an anisotropic continuum dressed by two lasers 
and a magnetic filed.  In the absence of the two lasers, the model reproduces the standard results of Fano-Feshbach resonances 
as can be  verified from  Eqs. (\ref{tmat}), (\ref{bme}), (\ref{ce}) and (\ref{fll}). In the absence of one laser and the magnetic field, 
the model reduces to that of $s$-wave \cite{fedichev}  and (or)  higher partial-wave OFR \cite{debprl}.  
In what follows we apply this theoretical method to two specific  atomic systems to show how  the excited-state 
bound-bound coherence described by the term $\mathcal{K}_{nn'}$  affects ground-state elastic and 
inelastic scattering properties.

For numerical illustrations, we consider two atomic systems: (A)   fermionic  $^6$Li 
alkali atoms  
and (B) spin-singlet ground-state bosonic $^{174}$Yb atoms.
We assume  that the two PA lasers $L_1$ and $L_2$ are co-propagating and linearly polarized in the $z$ axis
for both  systems. For this geometry, 
${\cal K}_{n n'} \ne  0$ only for  $m_{\ell} = M_{1} = M_{2}$  for both
 systems.

\subsection{Fermionic  $^6$Li atoms}

To select appropriate  excited states of $^6$Li$_2$ for our numerical illustrations, we 
consider low-lying rotational levels of a relatively deeply bound vibrational state of  
electronically excited 1$^{3}\Sigma_g^{+}$ potential which belongs to Hund's case b and 
asymptotically correlates to two separated  ($S_{1/2} + P_{1/2}$) atoms.  
We calculate several ro-vibrational states supported by this potential. The purpose is to  choose two  excited bound states 
which lie far below the dissociation threshold of the excited potential such that 
the possibility of laser transitions at the single-atom level can be safely ruled out.  
Furthermore,  rotational energy spacings  of such  bound states should be  large enough compared 
to other energy scales of our model. Hyperfine interaction in such excited bound states is assumed to 
be negligible compared to the resonant dipole-dipole interaction between the two atoms. 
The laser couplings $\Omega_{\chi n}$ and $\Lambda_{E,\ell m_{\ell}}^{J_n M_n}$
depend on the respective Franck-Condon (FC) overlap integral. 
To  estimate FC integrals, we  
consider the  $v=57$ vibrational state of the  1$^{3}\Sigma_g^{+}$ potential. 
This state 
has a vibrational energy 2575.3 GHz below the dissociation threshold and an outer turning point around 30 $a_0$.
 The rotational energy spacings
$\Delta E_{vJ} = E_{v,J+1} - E_{v, J}$ [where $E_{v,J}$ stands for the  binding energy of the  $(v,J)$ 
ro-vibrational state] are 
 3.4, 6.7  and 10.4 GHz for $J = 0 $, $J = 1$ and $J = 2$, respectively. 
The rotational  quantum number is given by 
$\vec{J}= \vec{N} = \vec{L} + \vec{\ell}$ where $L$ is the molecular electronic
orbital angular momentum.  The selection rule for laser transitions  dictates that $\Delta N = \pm 1$.

Exchange symmetry of a pair of spin polarized ground state fermionic $^6$Li atoms allows  
only odd partial waves in 
the ground continuum. In such a case, $p$-wave scattering state of the ground continuum can be optically coupled 
to both $J_{1}=0$ and $J_2=2$ rotational states by the two lasers. This can  lead to  the $p$-wave 
manipulation through optically generated coupling $\mathcal{K}_{n n'}$ between the two excited rotational 
states.

Let us now consider two-component  fermionic  $^6$Li 
alkali atoms initially prepared in hyperfine spin $f=1/2$ with  two magnetic components
 $m_{f} = - m_{f}' = 1/2$. In the presence of a strong  magnetic field $B$ (near $B \simeq B_0=834.1$ G for MFR), 
 neither atomic hyperfine quantum number $f$ nor its projection $m_f$
along the quantization axis is a good
quantum number, but the total magnetic quantum number $M_F = m_{f} + m_{f'}$ of the two atoms with hyperfine numbers $f$ and 
$f'$ remains  a good quantum number at any magnetic field strength.  In case of two-component $^6$Li atoms with  $M_F=0$, 
dipole selection rules  allow 
$d$-wave to be coupled  to both $J_1=1$ and $J_2=3$ rotational states by the two lasers while $s$-wave 
can be coupled to only $J_1=1$ rotational state. Henceforth, we focus only on $d$-wave manipulation. 

To estimate ground-state input parameters,  we calculate ground-state channel potentials in the presence of 
a magnetic field making use of the  singlet 
and triplet ground-state  potential data available in the literature \cite{pra:1995:hulet,jcp:1993:zumke}.
There are five  asymptotic channels for  $M_F=0$. We consider the lowest two asymptotic channels
as a two-channel model for our numerical work.  The method 
of calculation of ground-state channel potentials is described in some detail in  Appendix B. 
Since the closed channel 
quasibound state $\mid \chi \rangle $ has $s$-wave rotational angular momentum, by selection rules 
we have $\Omega_{\chi n} \ne 0 $ for $n=1$ and $\Omega_{\chi n} = 0$ for $n=2$. 
However, both the $s$-wave bare continuum and the  $\mid \chi \rangle $ 
state can be indirectly coupled to the $d$-wave via the $J=1$ bound state by
two-photon process with both photons 
coming from laser $L_1$  in the strong-coupling PA regime. Furthermore, 
the rotational selection rule dictates  $\Lambda_{00}^{J_2 M_2} = 0$ and
hence  ${\cal K}_{n n'} = {\cal V}_{n n'} -i{\cal G}_{n n'}/2$.
Since ${\cal V}_{n n'}$ is a small shift between the two excited states,
 for our numerical work we set ${\cal V}_{n n'} = 0$.  Thus we have ${\cal K}_{n n'} =  -i{\cal G}_{n n'}/2$ where
$ {\cal G}_{n n'} =   2 \pi \Lambda^{\ell=2 \, m_{\ell}}_{J_n M_n} \Lambda_{\ell=2 \, m_{\ell}}^{J_{n'} M_{n'}} $
implying that this coupling term arises  from the  dipole interactions of the two excited bound states with the $d$-wave 
ground scattering state only.

Let us suppose that the frequencies  of both  lasers are tuned  such that  $E - \hbar \delta_{n} - E_{n}^{\rm{shift}} \simeq 0$ 
for $n=1,2$. Then under the conditions mentioned  above, 
the the numerator on the RHS of Eq. (\ref{ae}) becomes real  provided the lasers do not introduce any  phase
as can be verified from  Eqs. (\ref{a0}) and (\ref{f1}). It then follows from  Eq. (\ref{inel})
that the terms that 
make positive contributions to  $\sigma_{\rm{inel}}$ are 
proportional to $(\mathscr{G}_1 + \gamma_1)/|{\cal D}_1|^2$  or 
$(\mathscr{G}_2 + \gamma_2)/|{\cal D}_2|^2$, where ${\cal D}_n = \mathscr{E}_n + i (\mathscr{G}_n + g_n)/2$. Therefore, in order 
to suppress inelastic scattering, the terms $\mathscr{G}_n$ have  to be reduced.   Now, since  
$\mathscr{G}_n = \Gamma_{n00}^{\rm{mf}} + \sum_{\ell >0, m_{\ell}} \Gamma_{n\ell m_{\ell}}
 +\Gamma_{n n'}$, and under the conditions
$\Gamma_{nn'}$ of Eq.(\ref{gnn'}) becomes equal to $ -  |{\cal G}_{n n'}|^2/(\Gamma_{n'} + \gamma_{n'})$ 
 implying suppression of inelastic scattering due to optically induced coherence between the two excited bound states. Furthermore, 
 since the $d$-wave ($\ell=2,m_{\ell}$) elastic cross section is proportional to the term 
$\sum_{\ell',  m_{\ell'}} |F_{\ell=2 m_{\ell}}^{\ell' m_{\ell'}}|^2$ which is basically the summation of  
amplitudes for  all the possible transitions from the two bound states to the $d$-wave ground continuum, the $d$-wave 
elastic scattering cross section would be enhanced when those transition amplitudes  add up coherently or 
in other words, interfere constructively.   The numerical
results discussed below further corroborate these analytical findings.

For $^6$Li atoms, the $s$-wave MFR is expected to influence the  $d$-wave scattering process at relatively low energy.
We find that for energies greater than
100 $\mu$K, the MFR has practically no influence.  Therefore, for higher
energies d-wave manipulation occurs mainly due to two lasers. This means
that for higher energies our method of d-wave manipulation is essentially
all optical.  This optical method can be 
 applicable for a wide range of energies in the sub millikelvin  regime. We express the  partial stimulated linewidth
 as $ \Gamma_{n, \ell m_{\ell}}   = g_{n} \mathscr{H}_{n,\ell m_{\ell}} $,  where $g_{n}$
is the dipole coupling proportional to the  intensity of the $n$th laser and  $ \mathscr{H}_{n,\ell m_{\ell}}$ is the angular part of the coupling.
We assume that $\gamma_1 = \gamma_2 = \gamma_{sp}$ and scale all energy quantities by $\hbar \gamma_{sp}$.
For all our numerical calculations, lasers $L_1$ and
$L_2$ are assumed to be on resonance with PA transitions between the ground continuum and the excited bound states 1 and 2, respectively.
 For $^6$Li, an energy-dependent $s$-wave background phase shift is estimated  from
the experimental parameters of the MFR  reported in Ref. \cite{revfesh} as discussed in  Appendix B. 
For $^6$Li atoms, 
the height  of the $d$-wave centrifugal barrier is about 50 mK and the peak of the
barrier is located at a separation of around 50$a_0$. 
Therefore,  the unperturbed $d$-wave scattering amplitude in the inner region of the barrier will be negligible
unless the initial collision energy is relatively higher.

In Fig.1, we show  $s$-wave and $d$-wave elastic and total inelastic scattering cross sections of $^6$Li  as a function of the intensity of the second laser  for  two different values of
the intensity ($g_1$) of the first laser at magnetic fields  $B$ near resonant value $B_0$  [ Figs. 1(a) and 1(c)]
and far away from the  MFR [Figs. 1(b) and 1(d)]. Comparing Fig. 1(a) with Fig. 1(b) and Fig. 1(c) with Fig. 1(d),
we infer that while the optical coupling of the  MFR
with the $d$-wave leads to significant enhancement of the $d$-wave scattering cross section in the strong-coupling regime, it has a marginal effect on inelastic scattering. It can also
be noticed that the intensity of the second laser has almost no effect on the  $s$-wave scattering amplitude for a fixed value of $B$. 
Comparing Fig.1(a) with Fig.1(c), we observe
that with the increase of the intensity of the first laser, the 
$d$-wave scattering cross section increases when $B$ is tuned near the MFR. 
Since $J_2 = 3$ rotational state is not accessible from the 
$s$-wave scattering state, the free-bound  stimulated linewidth ($\Gamma_n$) of the second ($n=2$) 
bound state is almost entirely composed of $d$-wave 
partial stimulated linewidth.  At 100 $\mu$K temperature, in order to to make 
$\Gamma_2 \simeq 1$ MHz,  the  intensity of laser $L_2$  has to  be  about 500 kW/cm$^2$, while to make 
$\Gamma_1 \simeq 1 $ MHz in microkelvin  temperature regime   the intensity of laser  $L_1$  is found to be of the order 
of 10 kW/cm$^2$. 
The unperturbed or background $d$-wave  phase shift at 100 $\mu$K temperature is found to be  smaller 
than  that of the $s$-wave background phase shift by four orders 
of magnitude. This means that the unperturbed $d$-wave scattering cross section
in the  microkelvin  temperature regime is extremely small.  Since the centrifugal barrier 
height is inversely proportional 
to the reduced  mass of the two atoms, for heavier atoms the $d$-wave barrier height is  lower. 
Therefore,  in case of heavier atoms it might be possible to manipulate the $d$-wave interaction more efficiently 
with much lower laser intensities as shown in the next section.  

\subsection{Bosonic $^{174}$Yb atoms }

The method of manipulation of interatomic interaction in  $^{174}$Yb atomic gas  is 
all optical because $^{174}$Yb has  no magnetic moment. 
For $d$-wave manipulation, as in the case of two-component $^6$Li atoms, it is necessary to tune the two  lasers 
$L_1$ and $L_2$  near the exited rotational levels $J_1=1$  and $J_2=3$, respectively. 
 For the excited Yb$_2$ molecule, $\vec{J}= \vec{J}_e + \vec{\ell}$ where $J_e$ 
 is the total electronic angular momentum.
The two excited states are assumed to belong to $0^{+}_{u}$  (Hund's case c) molecular symmetry, 
that is, the projection of $J_e$ on the internuclear axis is zero. The  $s$- and $d$-wave  background phase shifts and other 
input parameters for Yb  are estimated from the data reported in Refs. \cite{prl:2007:enomoto,pra:2009:takahashi:lineshape}.
The ground-state potential of Yb
is singlet only with no hyperfine structure and therefore the
ground-state scattering of Yb is essentially single-channel.

\begin{figure}
\includegraphics[width=3.5in]{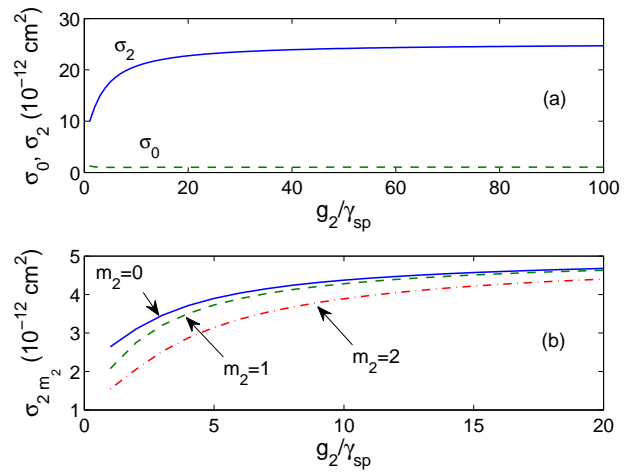}
\caption{These plots are for $^{174}$Yb atoms. (a) $\sigma_{0}$ (dashed) and  $\sigma_{2}$ (solid)  are plotted as a function of $g_2/\gamma_{sp}$.
(b) $\sigma_{2 m_{2}}$ Vs. $g_2/\gamma_{sp}$ for $m_2=0$ (solid), $m_2=1$ (dashed) and $m_2=2$ (dashed-dotted). The other parameters are the same as in Fig.2.}
\label{fig5}
\end{figure}

We plot total elastic and inelastic scattering
cross sections of $^{174}$Yb as a function of $g_2$ in Fig.2. From this figure, we notice that for extreme strong-coupling regimes inelastic scattering can be
suppressed by two  orders of magnitude as compared to elastic scattering. $g_2 = 200 \gamma_{sp}$ corresponds to 40 MHz. From the theoretical results reported
 in Ref. \cite{pra:2009:takahashi:lineshape}, we estimate that at a laser intensity of 1 W/cm$^2$,  the 
 $d$-wave stimulated 
 linewidth of $^{174}$Yb at 100 $\mu$K energy is about 0.1 MHz. Therefore, a 
40-MHz stimulated linewidth will correspond to laser intensity of  about 0.4 kW/cm$^2$.
Since  the spacing between the two excited bound states can be more than 50 MHz,
the possibility of exciting multiple states by the laser 1 or laser 2 is
unlikely. 

We compare $s$- and $d$-wave scattering cross sections in Fig. 3(a) which shows that the $d$-wave scattering cross section can exceed the $s$-wave one
by an  order of magnitude. Since  ${\cal K}_{n n'} \ne  0$ only for  $m_{\ell} = M_{1} = M_{2}$,
  ${\cal K}_{n n'}$ 
influences $d$-wave scattering for  $-1 \le m_{\ell=2} \le 1$ only while $m_{\ell} = \pm 2$ remains unaffected 
by this term. Figure 3(b)  demonstrates that  $\sigma_{2 m_{2}}$ for $m_2 = 0$ is higher compared  to those for $m_2=1$ and $m_2=2$
in the low  $g_2$ regime  but  $\sigma_{2 m_{2}}$ for different $m_2$ values tend to merge and saturate in the high $g_2$ regime. This means that in the low $g_2$ regime
the enhancement in $d$-wave scattering occurs predominantly  with an interatomic axis oriented along $z$ axis or the axis of quantization.

From Fig. 2  we estimate the reduced inelastic rate of $^{174}$Yb  for a  typical number density of $10^{13}$/cm$^3$. Taking
$\sigma_{inel} \sim 2 \times 10^{-13}$ cm$^2$ and the velocity at 100 $\mu$K temperature to be  $ \sim 10$ cm/s,  we find an inelastic rate
 $ \sim 20 $ s$^{-1}$. The elastic rate for $\sigma_{el} \sim 10^{-11}$ cm$^2$ would be two orders of magnitude larger than the inelastic rate.
These numbers indicate that optical manipulation of $d$-wave interaction in Yb is possible.

\section{conclusions } 
In conclusion, we have presented  an analytical  method for coherent manipulation of higher partial-wave atom-atom 
interaction by a pair of laser fields and an optional magnetic field in the strong photoassociative coupling regime. 
At low energy, this method is particularly applicable for manipulation of $p$- and $d$-waves. Although, 
we have discussed $p$-wave manipulation,  we have focused only on $d$-wave manipulation. The key effect  
predicted is the optically induced coherence between two rotational excitations. As a proof-of-principle, we 
have demonstrated the significance of this coherence in suppression of inelastic scattering and enhancement 
of $d$-wave elastic scattering. 
We have derived an analytical expression for the ground-state scattering $T$ matrix which clearly 
contains the effects of quantum interference.  
An analytical form of the two-body $T$ matrix is important as it may serve as a  
key ingredient for any theoretical treatment of interacting many-particle systems. In this work, we have 
discussed application of our method to two-component fermionic $^6$Li and bosonic  $^{174}$Yb atoms. However,
our method is applicable to all the atomic species that are currently being used in cold atom research. 
It would be particularly interesting to apply our method to explore
manipulation of  $p$- or  $d$-wave interaction in heavier  fermionic atoms such as 
$^{171}$Yb, $^{173}$Yb and $^{87}$Sr,  all of which have very narrow linewidths.  
 In order to make this  method widely applicable, it is also important to find ways of  reducing the  natural linewidth
of molecules  by either tailoring vacuum fluctuations  or by an interplay between light- and
vacuum-induced coherences \cite{daspra}.
 
 \acknowledgments{The author is thankful to C. Salomon,  R. Hulet and Saikat Ghosh  for discussions and comments.}

\appendix

\section{DERIVATION}

Here we present detailed derivation of the anisotropic dressed continuum  of Eq. (\ref{eq1}).
 From the time-independent Schr\"{o}dinger equation 
$\hat{H}\mid E,{\hat{k}} \rangle = E \mid E,{\hat{k}} \rangle$, making use of the expansions 
$B_E = \sum_{\ell' m_{\ell'}} B_{E}^{\ell' m_{\ell'}} Y_{\ell' m_{\ell'}}^{*}(\hat{k})$ and 
$A_{n E} = \sum_{\ell' m_{\ell'}} A_{n E}^{\ell' m_{\ell'}} Y_{\ell' m_{\ell'}}^{*}(\hat{k}) $, 
we obtain the following set of coupled algebraic equations
\bea 
(\hbar \delta_n - E) A_{nE}^{\ell'm_{\ell'}} + \Omega_{n\chi} B_{E}^{\ell'm_{\ell'}} = &-& \sum_{\ell m_{\ell}} \int dE' 
\Lambda_{J_n M_n}^{\ell m_{\ell}}(E') \nonumber \\
&\times& C_{E' \ell m_{\ell}}^{\ell'm_{\ell'}}(E), \label{cc1}
\eea
\bea 
(E_{\chi} - E) B_E^{\ell'm_{\ell'}} + \sum_{n,M_n} \Omega_{\chi n} A_{nE}^{\ell' m_{\ell'}} = &-& \int dE' V_{\chi 0}(E') \nonumber
\\ &\times& 
C_{E'00}^{\ell'm_{\ell'}}, \label{cc2}
\eea 
and
\bea 
(E' - E) C_{E'\ell m_{\ell}}^{\ell' m_{\ell'}}(E) &+& \sum_{n,M_n} \Lambda_{\ell m_{\ell}}^{J_n M_n}(E')
A_{nE}^{\ell' m_{\ell'}}
\nonumber \\ &+& V_{ 0 \chi} (E') B_{E}^{\ell' m_{\ell'}} \delta_{\ell 0} = 0 
\label{cc3} \eea
where $\delta_n = E_n/\hbar  -  (\omega_{L_n} - \omega_A) $ with
$E_n $ being the binding energy of $n$-th excited bound state measured from
 the threshold of the excited state potential,  
$\omega_{L_n}$ is the laser frequency of $n$-th laser and $\omega_A$ the atomic 
transition frequency. Since the coefficient $C_{E'\ell m_{\ell}}^{\ell' m_{\ell'}}$ is required 
to fulfill scattering boundary conditions at large separation of the two atoms, we can express Eq. (\ref{cc3}) in the form 
\bea 
C_{E' \ell m_{\ell}}^{\ell' m_{\ell'}}(E) &=& \delta (E-E') \delta_{\ell,\ell'} \delta_{\ell' m_{\ell'}} +
\frac{V_{0 \chi}(E')}{E-E'} B_{E}^{\ell' m_{\ell'}} \delta_{\ell,0} \nonumber \\ &+& 
\sum_{n,M_n} \frac{\Lambda_{\ell m_{\ell}}^{J_n M_n}(E')}{E - E'} A_{nE}^{\ell' m_{\ell'}} \label{ce1}
\eea
The partial-wave symbols appearing in the 
superscript and subscript of   $C_{E' \ell m_{\ell}}^{\ell' m_{\ell'}}(E)$ refer to the incident and scattered 
partial waves, respectively. 
Substituting Eq. (\ref{ce1}) in  Eqs. (\ref{cc1}) and (\ref{cc2}), we obtain 
\begin{widetext}
\bea 
(\delta_n - E) A_{nE}^{\ell'm_{\ell'}} + \Omega_{n \chi} B_{E}^{\ell'm_{\ell'}} &=& -  \Lambda_{J_n M_n}^{\ell' m_{\ell'}}(E)
- \left [ \int dE' \frac{V_{0 \chi}(E')  \Lambda_{J_n M_n}^{0 0}(E')}{E-E'} \right ] B_{E}^{\ell' m_{\ell'}}  - \sum_{\ell m_{\ell}}
\left [ \int dE' \frac{|\Lambda_{\ell m_{\ell}}^{J_n M_n}(E')|^2} {E - E'} \right ] 
  \nonumber \\&\times& A_{nE}^{\ell' m_{\ell'}}  -  \sum_{\ell m_{\ell}} \sum_{M_{n'}} \left [ \int dE' \frac{ \Lambda_{\ell m_{\ell}}^{J_n M_n}(E')
\Lambda^{\ell m_{\ell}}_{J_{n'} M_{n'}}(E') } {E - E'} \right ]  A_{n'E}^{\ell' m_{\ell'}},   \label{an}
\eea
\bea 
(E_{\chi} - E) B_E^{\ell'm_{\ell'}} + \sum_{n,M_n} \Omega_{\chi n} A_{nE}^{\ell' m_{\ell'}} = - V_{\chi 0}(E)\delta_{\ell' 0} - 
\int dE' \frac{|V_{\chi 0}(E')|^2}{E-E'} B_E^{\ell' m_{\ell'}} - \sum_{n,M_n} \int dE' \frac{V_{\chi 0}(E') \Lambda_{00}^{J_n M_n}}{E-E'} A_{nE}^{\ell' m_{\ell'}} 
\label{be} \eea 
\end{widetext}
where $n' \ne n$. In writing the third term on the right side of Eq. (A5) we have used the identity $ \sum_{M'} \langle J M \mid \hat{\epsilon}\cdot \hat{d} \mid \ell m_{\ell} \rangle \langle 
\ell m_{\ell} \mid \hat{\epsilon}\cdot \hat{d} \mid J M'\rangle = \sum_{M'} \langle J M \mid \hat{\epsilon}\cdot \hat{d} \mid \ell m_{\ell} \rangle \langle 
\ell m_{\ell} \mid \hat{\epsilon}\cdot \hat{d} \mid J M'\rangle \delta_{M,M'} = |\langle \ell m_{\ell}  \mid \hat{\epsilon}\cdot \hat{d} \mid  J M  \rangle|^2 $, where $\hat{\epsilon}$ stands for laser 
polarization and $\hat{d}$ is a unit vector pointing toward the  molecular dipole moment. Taking $E \rightarrow E + i\eta$ with $\eta=0^+$, we have 
\bea \int dE' \frac{V_{0\chi}(E')  \Lambda_{J_n M_n}^{0 0}(E')}{E-E'} &=& V_{n \chi}^{(\ell=0)}(E) 
- i \pi V_{0 \chi}(E) \nonumber \\
&\times& \Lambda_{J_n M_n}^{0 0}(E) \label{cb} \eea 
where 
\bea 
V_{n \chi}^{(\ell=0)}(E) = {\cal P}  \int dE'  \frac{ V_{0 \chi}(E')\Lambda_{J_n M_n}^{0 0}(E')}{E-E'}
\eea 
is an effective interaction between the closed-channel 
bound state $\mid \chi \rangle$ 
and the excited bound state $\mid b_n\rangle$ mediated through the  $s$-wave part of the ground  continuum. Similarly,  we have 
\bea 
\int dE' \frac{|V_{0\chi}(E')|^2}{E - E'} = E_{\chi}^{\rm{shift}} - i \frac{\Gamma_f}{2},
\label{chio}
\eea
\bea 
\int dE' \frac{|\Lambda_{\ell m_{\ell}}^{J_n M_n}(E')|^2  }{E - E'}
= E_{n\ell m_{\ell}}^{\rm{shift}} - i \frac{\Gamma_{n\ell m_{\ell}}}{2},
\label{nl}
\eea 
and
\bea
\int dE' \frac{ \Lambda_{\ell m_{\ell}}^{J_n M_n}(E') 
\Lambda^{\ell m_{\ell}}_{J_{n'} M_{n'}}(E') } {E - E'} = {\cal V}_{nn'}^{(\ell m_{\ell})} - i 
\frac{{\cal G}_{nn'}^{(\ell m_{\ell}}}{2} \nonumber \\
\label{sknn'} \eea
where  
\bea 
E_{\chi}^{\rm{shift}} = {\mathcal P}  \int dE' \frac{|V_{0\chi}(E')|^2}{E - E'}
\eea 
is the shift of the closed-channel bound state for its interaction with $s$-wave part 
of the open channel continuum, $\Gamma_f = 2 \pi |V_{0\chi}(E)|^2 $ is the magnetic Feshbach resonance line width, 
$ E_{n \ell m_{\ell}}^{\rm{shift}} = {\mathcal P} \int dE' \frac{|\Lambda_{\ell m_{\ell}}^{J_n M_n}(E')|^2 }{E - E'} $
is the partial light shift  and $\Gamma_{n\ell m_{\ell}}(E) = 2 \pi |\Lambda_{\ell m_{\ell}}^{J_n M_n}(E)|^2 $ is the partial stimulated line width of 
 the $n$-th excited bound state. Here 
\bea 
{\cal V}_{nn'}^{(\ell m_{\ell})} = {\mathcal P} \int dE' \frac{ \Lambda_{\ell m_{\ell}}^{J_n M_n}(E') 
\Lambda^{\ell m_{\ell}}_{J_{n'} M_{n'}}(E') } {E - E'}, \eea  
\bea 
{\cal G}_{nn'}^{(\ell m_{\ell})} = 2 \pi \Lambda_{\ell m_{\ell}}^{J_n M_n}(E) 
\Lambda^{\ell m_{\ell}}_{J_{n'} M_{n'}}(E). \eea 

Substituting Eqs.(\ref{cb}), (\ref{chio}), (\ref{nl}) and (\ref{sknn'}) into  Eqs. (\ref{an}) and (\ref{be}),
we obtain 
\begin{widetext}
\bea 
(E - \delta_n - E_n^{\rm{shift}} + i \Gamma_{n}/2 ) A_{nE}^{\ell'm_{\ell'}}  - \{ \Omega_{n\chi} + V_{n\chi}^{(\ell=0)} - i\pi \Lambda_{J_n M_n}^{00} V_{0\chi}(E) \}  B_{E}^{\ell'm_{\ell'}} = 
 \Lambda_{J_n M_n}^{\ell' m_{\ell'}}(E) + \sum_{M_{n'}} {\cal K}_{n n'}^{\rm{LL}} A_{n'E}^{\ell' m_{\ell'}} \label{ann}
\eea
\bea 
(E - E_{\chi} - E_{\chi}^{\rm{shift}} + i\Gamma_f/2 ) B_E^{\ell'm_{\ell'}} - \sum_{n,M_n} \{ \Omega_{\chi n} + V_{n\chi}^{(\ell=0)} - i \pi \Lambda^{J_n M_n}_{00} V_{\chi 0}(E) \} A_{nE}^{\ell' m_{\ell'}}
 =  V_{\chi 0}(E) \delta_{\ell' 0}
\label{ben}  
\eea
\end{widetext}
where $E_n^{\rm{shift}} = \sum_{\ell m_{\ell}} E_{n \ell m_{\ell}}^{\rm{shift}}$ is the total light shift due to laser 
$L_n$, $\Gamma_n= \sum_{\ell m_{\ell}} \Gamma_{n\ell m_{\ell}}$
is the corresponding total stimulated linewidth and
\bea 
{\cal K}_{nn'}^{\rm{LL}}  = \sum_{\ell m_{\ell}}\left  ({\cal V}_{nn'}^{(\ell m_{\ell})} - i \frac{1}{2} {\cal G}_{nn'}^{(\ell m_{\ell}} \right )
\label{knnpll}
\eea 
is  an effective complex coupling 
between the two excited  bound states $\mid b_n \rangle $ and $\mid b_{n'\ne n} \rangle $ induced by the two 
lasers through  optical dipole interactions.  The superscript LL is used to indicate the use of two lasers. 
Equation (\ref{ben})  can be expressed in a compact form 
\bea 
B_E^{\ell'm_{\ell'}} &=&  \frac{2}{(\epsilon + i) \Gamma_f } \left 
[   V_{\chi 0}(E) \delta_{\ell' 0} + \sum_{n,M_n}  (q_{nf} - i) \right. \nonumber \\
&\times& \left.   \pi \Lambda^{J_n M_n}_{00} V_{\chi 0}(E) 
A_{nE}^{\ell' m_{\ell'}} \right ] \label{beq}
\eea
where 
\bea 
\epsilon =  \frac{E - E_{\chi} - E_{\chi}^{\rm{shift}}}{\Gamma_f/2} \label{epsn} 
\eea
is the dimensionless collision energy measured from the shifted binding energy $E_{\chi} + E_{\chi}^{\rm{shift}}$ 
of the closed channel quasibound state $\mid \chi \rangle $ and 
\bea
q_{nf}= \frac{\Omega_{\chi n} + V_{n\chi}^{(\ell=0)}}{\pi \Lambda^{J_n M_n}_{00} V_{\chi 0}(E)}
\eea
is the Fano-Feshbach asymmetry parameter that describes quantum interference between two transition pathways 
(i) $\mid b_n \rangle \xrightarrow{L_n\,\rm{Laser}} \mid \chi \rangle$ and 
(ii) $ \mid b_n \rangle \xrightarrow{L_n\,\rm{Laser}} \mid E \ell=0 \, m_{\ell}=0 \rangle_{\rm{bare}} 
\xrightarrow{\rm{hf}}\mid \chi \rangle$. Substituting Eq. (\ref{beq}) into Eq. (\ref{ann}), we have 
\bea 
&& \left (E - \delta_n - E_n^{\rm{shift}} + i \frac{\Gamma_{n}}{2} 
- \frac{(q_{nf} - i)^2}{\epsilon +i} \frac{\Gamma_{n00}}{2} \right ) A_{nE}^{\ell'm_{\ell'}} \nonumber \\
&-&  \sum_{M_{n'}} {\cal K}_{nn'}(M_n, M_{n'}) A_{n'E}^{\ell' m_{\ell'}}(M_{n'}) \nonumber \\
&=& \frac{q_{nf} - i}{\epsilon + i} \Lambda_{J_n M_n}^{00}(E)\delta_{\ell' 0}  + 
 \Lambda_{J_n M_n}^{\ell' m_{\ell'}}(E). 
\eea
Here 
\bea
{\cal K}_{nn'}(M_n, M_{n'}) =  {\cal K}_{nn'}^{\rm{mf}}  + {\cal K}_{nn'}^{\rm{LL}} \label{knnp} \eea  
where 
\bea
{\cal K}_{nn'}^{\rm{mf}} =
(q_{nf} -i)(q_{n'f} - i) \pi \Lambda_{J_n M_n}^{00}  \Lambda^{J_{n' } M_{n'}}_{00}.
\label{knnpmf}
\eea  
Note that, when both the lasers are linearly polarized along the quantization axis, we have 
${\cal K}_{nn'}(M_n, M_{n'}) = 0$ unless $M_n = M_{n'}$. 
Putting  $n=1,2$ and $n'=1,2$ with $n'\ne n$ in Eq.(A17), one can write down  two coupled equations which 
can be solved to obtain 
\bea
 A_{nE}^{\ell'm_{\ell'}}(M_n) =
 \frac{\mathscr{A}_{n0}\delta_{\ell',0} + \mathscr{F}_{n\ell'} (1 - \delta_{\ell' 0}) }{{\cal D}_n} \label{ane} \eea 
 where 
\bea 
\mathscr{A}_{n0} =  \beta_{n \epsilon} \Lambda^{0 0}_{J_n M_n} +
\frac{1} {\xi}_{n'}  {\cal K}_{nn'} \beta_{n'\epsilon} \Lambda^{0 0}_{J_{n'} M_{n'}},   \label{a0}
\eea
\bea 
\mathscr{F}_{n\ell'} =
 \Lambda^{\ell' m_{\ell'}}_{J_n M_n}  + \frac{1} {\xi}_{n'} {\cal K}_{nn'} 
 \Lambda^{\ell' m_{\ell'}}_{J_{n'} M_{n'}}   \label{f1} \eea
 and 
 \bea 
 {\cal D}_n(E) = \xi_n  &-& \frac{(q_{nf} - i)^2}{ \epsilon + i } \frac{\hbar\Gamma_{n00}}{2} \nonumber \\
&-&   {\xi}_{n'}^{-1}{\cal K}_{nn'}{\cal K}_{n'n}.  \label{de}
\eea 
In Eq. (\ref{a0}),  $ \beta_{n\epsilon}$ is a dimensionless factor defined by  
  \bea \beta_{n\epsilon} = \frac{ q_{nf} + \epsilon}{\epsilon + i} \label{beta} \eea 
and in Eq. (\ref{de}) $ \xi_n(E)$ is given by 
 \bea  \xi_n(E) = E - \hbar \delta_{n} - E_{n}^{\rm{shift}} + i \hbar [\gamma_{n} 
+ \Gamma_{n}(E)]/2 \eea   
 In the above equations $n' \ne  n$ and $M_n = M_{n'}$.
We have here phenomenologically introduced the spontaneous linewidth  $\gamma_n$ by making the replacement 
 $\Gamma_n \rightarrow \Gamma_n + \gamma_n$.    
  Writing Eq. (\ref{de}) in the form   
  \bea {\cal D}_n(E) = \mathscr{E}_n + i \hbar (\mathscr{G}_n + \gamma_n)/2, 
  \eea 
 we have 
 \bea 
 \mathscr{E}_n = E - \hbar \delta_{n} - E_{n}^{\rm{shift}} - \mathscr{E}_{n}^{\rm{mf}} - \mathscr{E}_{n n'} 
 \eea 
 \bea 
 \mathscr{G}_n = \Gamma_{n00}^{\rm{mf}} + \sum_{\ell >0, m_{\ell}} \Gamma_{n\ell m_{\ell}}
 +\Gamma_{n n'}
 \eea 
where 
\bea
\mathscr{E}_{n}^{\rm{mf}} = \left [ \frac{(q_{nf}^2-1)\epsilon - 2 q_{nf}}{\epsilon^2 +1} \right ] 
\frac{\Gamma_{n00}}{2} \label{shiftmf}
\eea
\bea
\mathscr{E}_{n n'} = {\rm Re} [\xi_{n'}^{-1} \mathcal{K}_{n n'} \mathcal{K}_{n' n} ] \label{shiftLL}
\eea
\bea
\Gamma_{n00}^{\rm{mf}} = |\beta_{n\epsilon}|^2 \Gamma_{n00} 
\eea
\bea
\Gamma_{n n'} = - \frac{2}{\hbar}  \Im [\xi_{n'}^{-1} \mathcal{K}_{n n'} \mathcal{K}_{n' n} ] \label{gnn'}
\eea

\section{GROUND STATE OF TWO-COMPONENT  $^6$Li ATOMS  IN  A  MAGNETIC FIELD}

In the presence of a magnetic field,  the angular basis states of the form $\mid {f f', m_f, m_{f'}}, M_F \rangle$ (diabatic basis) 
of two separated atoms  for different values of $f,f',m_f$ and $m_{f'}$ get mixed up. 
The Hamiltonian of two ground state atoms at large separation  in the presence of an external magnetic field is then 
diagonalizable in these diabatic basis and the  eigenfunctions are regarded as the scattering channels. 
However, these are not diagonal at short separations where the Hamiltonian 
is diagonalizable in the molecular adiabatic basis. In the absence of hyperfine interactions, 
there are two adiabatic  molecular potentials corresponding to the  spin singlet and triplet, both of which asymptotically 
go as  $\sim - 1/r^6$.  The diabatic and adiabatic basis are related via transformation involving Wigner  9-$j$ symbol. For 
$M_F=0$ there are five  channels which become coupled at intermediate separations ( $\le 30a_0$, where $a_0$ is Bohr radius). 
The coupling is most significant at separations around 20 $a_0$. 
We consider the two lowest asymptotic channels 
with $M_F=0$. The lower channel $\mid g_1 \rangle $ is open and the upper channel $\mid g_2 \rangle$ is closed.  The 
asymptotic energy spacing between these two channels at $B_0$ is calculated to be 2.25 GHz. 
The open channel is mostly electronic  spin triplet 
(about 90\% triplet) while the closed channel is largely spin singlet. 
Exchange symmetry of a pair of two-component fermionic atoms with $M_F=0$ allows only even partial waves 
of relative motion of the pair, while for a pair of single component fermionic atoms only odd partial waves are allowed. 
For $^6$Li atomic gas initially prepared in a 50:50 mixture of two hyperfine components of $f=1/2$ and $M_F=0$, even partial-wave  
atom-atom interactions  are expected to be predominant. 

At low energy,  the Feshbach resonance width $\Gamma_f$ is given by
$\Gamma_f/2 \simeq k a_{bg} \Gamma_0$, where $k$ is the wave number related to the collision energy $E=\hbar^2 k^2/(2 m)$ with $m$ being the reduced mass of the two atoms, 
 $a_{bg} = -1405 a_0$ is the back ground scattering length, $\Gamma_0$ is a parameter related to the width $\Delta = -300$  G of the zero crossing by   
 $\Delta = \frac{\Gamma_0}{\delta \mu}$, with  $\delta \mu = 2 \mu_B $ (where $\mu_B$ is the Bohr magneton) being the difference of magnetic moment of closed-channel quasibound state 
from that of the two free atoms in the open channel. The energy-dependence of the Feshbach resonance phase shift $\eta_{res}$ is given by 
 $ - \cot \eta_{res} = \epsilon = \frac{E - E_{th}}{\Gamma_f/2} - \frac{E_{c} - E_{th}}{\Gamma_f/2}$, where $E_{th}$ is the threshold of the open channel that depends on the applied
magnetic field, and 
$ E_{c} - E_{th} = \delta \mu (B - B_0)$.  The background $s$-wave phase shift (with effective range of 
background open channel potential assumed to be zero)   $\eta_0^{bg}$ is given by 
$\cot \eta_0^{bg} \simeq - \frac{1}{k a_{bg}}$.
Sine the  coupling between open and closed channels occurs at an intermediate  separation ($k r \ll 1$), we    
can write $\Gamma_f/2 = \pi |V_{\chi 0} |^2 =  \pi | \int dr  \psi_\chi(r)  V(r)  \psi_{E, \ell=0}(r) |^2 \simeq 
\frac{k a_{bg}}{ 1 + (k a_{bg})^2} \Gamma_0$, where $V (r)$ stands for coupling 
between the two channels and $\psi_{E, \ell=0}(r) = \langle r \mid E, \ell=0 \rangle_{bare}$ is 
the energy normalized background (open channel) scattering state, which asymptotically behaves 
as $\psi_{E,00}(r \rightarrow \infty) \sim \sin(k r + \eta_0^{bg}) $ and   $\psi_\chi(r) = \langle r \mid \chi \rangle$

\end{document}